\documentclass[aps,prd,reprint,groupedaddress,amsmath,amssymb,floatfix]{revtex4-2}

\usepackage{graphicx} 
\usepackage{dcolumn}  
\usepackage{bm}       
\usepackage{capt-of}  
\usepackage{mathrsfs} 
\usepackage[colorlinks=true,allcolors=blue]{hyperref} 
\usepackage{xcolor}
\newcommand{\pt}{p_T}
\newcommand{\Ds}{D_s^+}
\newcommand{\Dz}{D^0}
\newcommand{\vtwo}{v_2}
\newcommand{\qtwo}{q_2}
\newcommand{\Tc}{T_c}

\begin{document}

\title{$D^0$–$D_s^+$ Elliptic-Flow Splitting under Event-Shape Engineering: A Probe of Sequential Charm Hadronization}
 
\author{Yu-Jie Huang}
\affiliation{School of Mathematics and Physics, China University of Geosciences, Wuhan 430074, China}

\author{Wei Dai}
\email{corresponding author: weidai@cug.edu.cn} 
\affiliation{School of Mathematics and Physics, China University of Geosciences, Wuhan 430074, China}

\author{Jiaxing Zhao}
\affiliation{Helmholtz Research Academy Hesse for FAIR (HFHF), GSI Helmholtz Center for Heavy Ion Physics, Campus Frankfurt, 60438 Frankfurt, Germany}
\affiliation{Institut f\"ur Theoretische Physik, Johann Wolfgang Goethe-Universität, Max-von-Laue-Straße 1, D-60438 Frankfurt am Main, Germany}

\author{Tan Luo}
\affiliation{School of Physics and Electronics, Hunan University, Changsha 410082, China}

\author{Ben-Wei Zhang}
\affiliation{Key Laboratory of Quark \& Lepton Physics (MOE) and Institute of Particle Physics, Central China Normal University, Wuhan 430079, China}

\author{Enke Wang}
\affiliation{State Key Laboratory of Nuclear Physics and Technology, Institute of Quantum Matter, South China Normal University, Guangzhou 510006, China}

\affiliation{Guangdong Basic Research Center of Excellence for Structure and Fundamental Interactions of Matter, Guangdong Provincial Key Laboratory of Nuclear Science, Guangzhou 510006, China}

\date{\today}

\begin{abstract}
Recent work has proposed sequential hadronization of open-charm hadrons in the quark-gluon plasma, wherein more tightly bound species such as $\Ds$ form earlier near $1.2\Tc$ and $\Dz$ forms later at $\Tc$. That work showed that this mechanism naturally reverses the sign of the $\Dz-\Ds$ elliptic-flow splitting $\Delta\vtwo\equiv\vtwo(\Dz)-\vtwo(\Ds)$ relative to the conventional simultaneous baseline. In this work, we demonstrate that event-shape engineering (ESE) provides a sharper discrimination between the two pictures than inclusive measurements alone. By selecting large-$\qtwo$ and small-$\qtwo$ events in 0--10\% and 30--50\% centrality classes in Pb--Pb collisions at $\sqrt{s_{\mathrm{NN}}}=5.02$ TeV, we show that the geometry-driven enhancement of charm-meson $\vtwo$ can be separated from the hadronization-time response: the positive $\Delta v_2(\Dz-\Ds)$ at intermediate $\pt$ in the sequential scenario grows systematically with $\qtwo$, while the corresponding response slope $\chi$ reveals a species-dependent hierarchy $\chi(\Dz) > \chi(\Ds)$ that is robust against the overall flow normalization and absent in the simultaneous baseline. In the simultaneous case, the splitting is near zero or negative and does not follow the same geometry scaling. Notably, the semi-central 30--50\% class emerges as the optimal window, because the non-monotonic interplay between QGP lifetime and initial eccentricity maximizes the late-stage flow conversion. The $\qtwo$ ratios of $\Ds/\Dz$ yield ratio remain close to unity, supporting that the splitting is a dynamical flow effect rather than a chemical yield modification. These results suggest $\Delta\vtwo(\Dz-\Ds)$ and the response slope $\chi$ under ESE as complementary differential probes of the space-time structure of charm hadronization near the QCD transition temperature.
\end{abstract}

\maketitle

\section{Introduction}

Heavy quarks are among the most informative probes of the strongly interacting matter created in relativistic heavy-ion collisions. Because charm and bottom quarks are produced predominantly in the earliest hard partonic scatterings, their total yields are largely fixed before the formation of the quark-gluon plasma (QGP). During the subsequent evolution, heavy quarks interact with the expanding medium, exchange energy and momentum with thermal partons, and partially inherit the collective flow of the QGP. The final spectra and azimuthal anisotropies of open heavy-flavor hadrons therefore encode both the transport history of heavy quarks in the deconfined medium and their hadronization mechanism near the QCD crossover~\cite{Rapp:2018qla,Dong:2019}.

The hadronization of charm quarks is especially important at low and intermediate transverse momentum. In addition to vacuum-like fragmentation, charm quarks may combine with nearby thermal light or strange quarks through coalescence. Such recombination naturally enhances charm-hadron species whose light-flavor content is abundant in the medium, modifies the baryon-to-meson and strange-to-nonstrange charm-hadron ratios, and transfers part of the collective flow of the light partons to the final heavy-flavor hadrons~\cite{Greco:2003mm,Fries:2003kq,Fries:2008hs}. Measurements of $\Dz$, $\Ds$, and other open-charm hadrons have therefore become central tools for constraining the microscopic degrees of freedom and flavor composition of the hadronizing medium.

Most coalescence calculations assume that different charm-hadron species are formed on a common hadronization hypersurface~\cite{Fries:2003kq,Greco:2003mm,Song:2015sfa,Plumari:2017ntm,Beraudo:2017gxw,He:2019tik,Cao:2019iqs,Andronic:2021erx,Zhao:2024ecc}. This simultaneous-hadronization assumption simplifies calculations, but it neglects the possibility that hadrons with different binding energies or light-flavor content may become stable at different temperatures as the QGP cools through the crossover region~\cite{Nahrgang:2013xaa,Bellwied:2013cta}. Sequential hadronization, inspired by the well-established formation and dissociation of quarkonia at different temperatures in the QGP, offers a more differential picture: different open-charm species can leave the partonic phase on distinct hypersurfaces.

Because charm-quark number is conserved, the earlier formation of one species reduces the pool available to later-forming hadrons and changes both the chemical yields and the dynamical flow of the remaining charm quarks. If the more tightly bound $\Ds$ meson can hadronize earlier than $\Dz$, the $\Ds/\Dz$ ratio and, more importantly, the relative elliptic flow between the two species become direct probes of the hadronization-time ordering.

In the simultaneous-hadronization baseline, $\Dz$ and $\Ds$ mesons are produced on the same $\Tc$ hypersurface. Coalescence with thermal strange quarks can favor a larger partonic flow for the parent charm quarks of $\Ds$, which tends to produce $\vtwo(\Ds)\gtrsim\vtwo(\Dz)$ at low and intermediate $\pt$---a trend expected in transport-plus-coalescence calculations~\cite{Cao:2015hia,Plumari:2017ntm,Beraudo:2014boa,He:2019vgs,Zhao:2023nrz}. In the sequential picture, however, the earlier formation of $\Ds$ near $1.2\Tc$ removes its parent charm quarks from the partonic phase sooner, while the charm quarks destined to become $\Dz$ remain coupled to the medium until $\Tc$ and accumulate additional elliptic flow during the interval $1.2\Tc \to \Tc$. This mechanism, introduced in a recent letter that established the sequential-hadronization framework for open heavy flavors~\cite{Xu:2025ivv}, produces a sign reversal of the flow difference, yielding a positive hadronic splitting
\begin{equation}
\Delta \vtwo(\Dz-\Ds) \equiv \vtwo(\Dz)-\vtwo(\Ds).
\end{equation}

The contrasting space-time pictures of simultaneous and sequential hadronization are illustrated schematically in Fig.~\ref{fig:schematic}.

The $\Delta\vtwo(\Dz-\Ds)$ measured without event-shape selection already discriminates between the two hadronization pictures. Event-shape engineering (ESE), previously employed by ALICE to measure D-meson $\vtwo$ as a function of $\qtwo$~\cite{ALICE:2018gif,ALICE:2020iug} and studied theoretically in the Catania transport framework~\cite{Sambataro:2022sns}, provides a sharper test when applied to the sequential-versus-simultaneous question. The central methodological challenge in isolating hadronization-time effects is that the elliptic flow of heavy-flavor hadrons depends jointly on the initial collision geometry and the duration of partonic evolution. ESE addresses this by providing an independent handle on geometry within a fixed centrality class, enabling a differential measurement of the hadronization-time response. This two-knob framework---one knob for the geometry ($\qtwo$) and one for the hadronization time ($T_{\rm form}$)---is not limited to open charm; it can be generalized to bottom hadrons, quarkonia, or any other species whose formation temperature differs from the bulk freeze-out temperature.

Within a fixed centrality interval, ESE selects large-$\qtwo$ and small-$\qtwo$ events that correspond to stronger and weaker initial elliptic geometry, respectively. The key question is how the $\Dz-\Ds$ splitting responds: if it were controlled only by global geometry, it should follow the same large-$\qtwo$/small-$\qtwo$ ordering as the individual charm-meson $\vtwo$; if it is controlled by sequential hadronization, it should additionally reflect the finite hadronization-time interval between $1.2\Tc$ and $\Tc$ and exhibit a characteristic $\qtwo$ dependence absent in the simultaneous baseline. Using ESE to isolate the hadronization-time response from the bulk geometry response is the central advance of the present work.
\begin{center}
    \centering
    \includegraphics[width=1.05\linewidth]{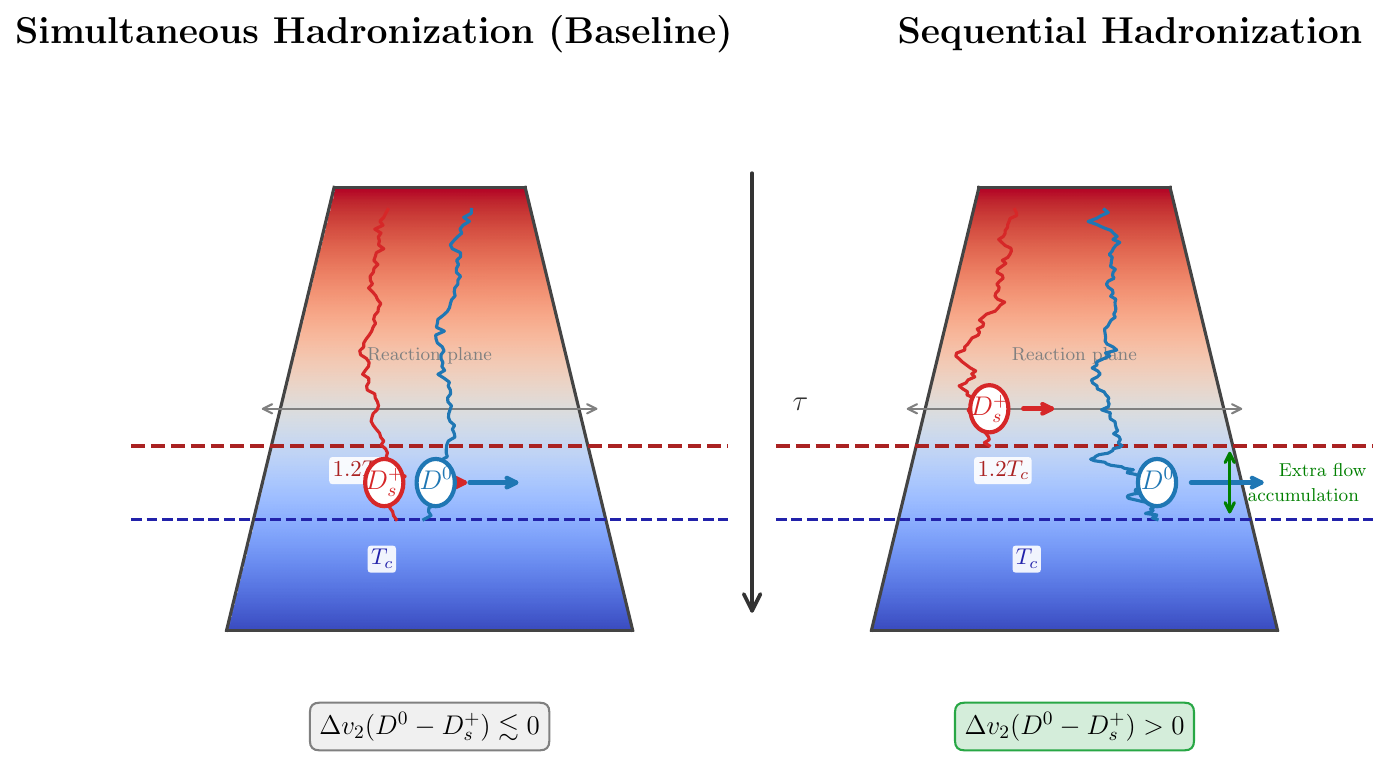}
    \captionof{figure}{Schematic illustration of simultaneous (left) and sequential (right) hadronization of open-charm mesons in an expanding QGP fireball. In the simultaneous baseline, $\Dz$ and $\Ds$ form at the same $\Tc$ hypersurface and accumulate similar elliptic flow. In the sequential picture, $\Ds$ freezes out earlier near $1.2\Tc$, while the parent charm quark of $\Dz$ remains interact with the medium until $\Tc$ and accumulates additional $\vtwo$ during the interval $1.2\Tc \to \Tc$, leading to a positive $\Delta \vtwo(\Dz-\Ds)$.}
    \label{fig:schematic}
\end{center}
In this work, we extend the sequential-hadronization framework of Ref.~\cite{Xu:2025ivv} to include event-shape engineering and present a comprehensive study in Pb--Pb collisions at $\sqrt{s_{\mathrm{NN}}}=5.02$ TeV. Comparing 0--10\% and 30--50\% centrality classes, we examine the $\pt$-differential elliptic flow, the $\Dz-\Ds$ flow splitting and its $\qtwo$ dependence, the $\Ds/\Dz$ chemistry and coalescence fraction, and the hadronization-time difference, to establish a geometry-resolved discrimination between the two hadronization pictures. The paper is organized as follows. Section~\ref{sec:model} describes the event-shape selection, charm transport, and hadronization framework. Section~\ref{sec:results} presents the flow splitting and hadron-chemistry observables. Section~\ref{sec:summary} summarizes the implications for sequential hadronization.

\section{Event-shape selected Langevin framework}
\label{sec:model}

Our calculations are performed within the updated SHELL framework introduced in Ref.~\cite{Xu:2025ivv}, which couples event-by-event (3+1)D viscous hydrodynamics for the QGP medium, Langevin transport of charm quarks, and a hybrid coalescence-plus-fragmentation hadronization model. Crucially, we employ the same event-by-event medium background and the same Langevin transport parameters for all charm-hadron species in both the sequential and simultaneous scenarios. The difference between the two hadronization pictures is introduced only at the final hadronization stage, ensuring that any species-dependent differences can be traced, within the present setup, to the ordering of charm-hadron formation.

The initial entropy density of each Pb--Pb collision event is generated from an event-by-event initial condition model and evolved with the (3+1)D CLVisc hydrodynamic code~\cite{Pang:2018zzo,Wu:2021fjf}, which provides the local temperature $T(x)$ and four-velocity $u^\mu(x)$ of the QGP. The hydrodynamic evolution determines both the transport of charm quarks and the properties of thermal light and strange quarks used in the coalescence calculation. Centrality classes are defined according to the final charged-particle multiplicity, and the same centrality selection is used when constructing the ESE classes.

Charm quarks are produced in primordial hard scatterings. Their initial momentum distribution is calculated with the FONLL approach~\cite{Cacciari:2012ny}, and their transverse positions are sampled according to the binary-collision density $n_{\rm coll}(\mathbf{r}_\perp)\propto T_A(\mathbf{r}_\perp)T_B(\mathbf{r}_\perp-\mathbf{b})$ of the MC-Glauber model~\cite{Miller:2007ri}. Within a fixed centrality interval, the average initial charm-quark production profile does not differ significantly between large-$q_2$ and small-$q_2$ events; any observed species-dependent flow differences therefore arise dominantly from the subsequent medium evolution and hadronization-stage ordering.

The event-shape selection isolates events with different initial eccentricities while keeping the average multiplicity within a narrow centrality window. Both the hydrodynamic evolution and the heavy-quark transport are carried out event by event, so that each collision retains its own initial-state fluctuation and medium history; the $q_2$ selection is therefore applied to an ensemble of individually evolved events. It is based on the reduced second-order flow vector $q_2=|Q_2|/\sqrt{M}$, where $Q_2$ is the second-harmonic flow vector constructed from the azimuthal angles of charged particles and $M$ is the multiplicity. We define the largest-$q_2$ and smallest-$q_2$ event classes as the top and bottom 20\% of the $q_2$ distribution, respectively.

The heavy-quark dynamics is described by a Langevin transport approach in the hydrodynamically expanding medium~\cite{Wang:2019xey,Dai:2018mhw,Wang:2020ukj,Wang:2020qwe,Cao:2013ita,Gossiaux:2009mk}. For a charm quark with energy $E=\sqrt{p^2+M^2}$, the position and momentum updates are written schematically as
\begin{align}
dx_i &= \frac{p_i}{E}dt,\\
dp_i &= -\Gamma(p)p_i dt + \sqrt{\kappa(p)dt}\rho_i - dp_i^{\mathrm{rad}},
\end{align}
where $\Gamma$ is the drag coefficient, $\kappa$ is the momentum diffusion coefficient, $\rho_i$ is a Gaussian stochastic force satisfying $\langle \rho_i(t)\rho_j(t')\rangle = \kappa\delta_{ij}\delta(t-t')$, and $dp_i^{\mathrm{rad}}$ represents recoil from medium-induced gluon radiation. The drag and diffusion coefficients are related through the fluctuation-dissipation relation $\kappa=2\Gamma ET$~\cite{Kubo:1966fyg} and are conventionally expressed in terms of the spatial diffusion coefficient $\mathcal{D}_s=2T^2/\kappa$. We adopt a constant $2\pi T \mathcal{D}_s=2.5$, consistent with recent (2+1)-flavor lattice-QCD calculations~\cite{Altenkort:2023oms} and previous Bayesian constraints from D-meson data~\cite{Xue:2025hrw}. The Higher-Twist approach~\cite{Guo:2000nz,Zhang:2003wk,Zhang:2004qm,Majumder:2009ge} is employed to simulate the medium-induced gluon radiation. The differential gluon radiation spectrum takes the form
\begin{equation}
\frac{dN_g}{dx dk_\perp^2 dt} = \frac{2\alpha_s C_s P(x)\hat{q}}{\pi k_\perp^4} \sin^2\!\left(\frac{t-t_i}{2\tau_f}\right) \left[\frac{k_\perp^2}{k_\perp^2+(xM)^2}\right]^4,
\end{equation}
where $x$ and $k_\perp$ are the energy fraction and transverse momentum of the emitted gluon, $C_s$ is the quadratic Casimir in color representation, $P(x)$ is the vacuum splitting function~\cite{Deng:2009ncl}, and $\tau_f=2Ex(1-x)/[k_\perp^2+(xM)^2]$ is the formation time. The jet transport parameter is parametrized as $\hat{q}(\tau,r)=\hat{q}_0(T/T_0)^3(p_\mu u^\mu)/p^0$ with $\hat{q}_0=1.5$~GeV$^2$/fm~\cite{Xie:2019oxg}. The temperature dependence of the heavy-quark spatial diffusion coefficient has been studied in our recent Bayesian analysis within the simultaneous improved Langevin framework ~\cite{Xue:2025hrw}. In the present work, we keep the transport setup fixed in order to isolate the effect of sequential hadronization. A combined study of sequential hadronization with temperature- and momentum-dependent transport coefficients will be investigated separately.

Hadronization is implemented through a hybrid coalescence-plus-fragmentation model~\cite{Greco:2003mm,Fries:2003kq,Ravagli:2007xx,Fries:2008hs}. Charm quarks first attempt to combine with thermal light or strange quarks from the medium. The coalescence probability depends on the overlap between the charm-quark phase-space distribution and the Wigner function of the final hadron. For a charm meson $M(c\bar{q})$, this probability can be written as
\begin{align}
\label{eq:coal}
\frac{dN_M}{d^3{\bf P}} &= g_M \int \prod_{i=1}^n\frac{d^3x_i\,d^3p_i}{(2\pi)^3E_i} f_i({\bf x}_i,{\bf p}_i) \\
&\quad \times W_M({\bf x}_1,\cdots,{\bf x}_n, {\bf p}_1,\cdots, {\bf p}_n)\, \delta^{(3)}\!\left({\bf P}-\sum_{i=1}^k{\bf p}_i\right), \nonumber
\end{align}
where $g_M$ is the statistical degeneracy factor, $n=2$ for mesons and $3$ for baryons, and $f_i(\mathbf{x}_i,\mathbf{p}_i)$ is the phase-space distribution of constituent quark $i$. Light quarks are assumed to be thermalized with the distribution $f_q = g/(e^{u_\mu p^\mu/T}+1)$. The Wigner functions for charmed mesons, including the ground and low-lying excited states ($1S$--$2P$), are approximated by harmonic oscillators with widths determined by the corresponding average radii~\cite{Zhao:2018jlw,Zhao:2020rrk,Zhao:2023nrz,Zhao:2025cnp}; explicit expressions and the definitions of relative coordinates can be found in Ref.~\cite{Xu:2025ivv}. The relative momentum of the two constituent quarks in their center-of-mass frame is defined as $\mathbf{p}=(m_q\mathbf{p}_c-m_c\mathbf{p}_q)/(m_q+m_c)$, and the Gaussian width $\sigma$ is fixed by the average radius $\langle r\rangle = \int d^3r d^3p/(2\pi)^3\, W_h(\mathbf{r},\mathbf{p})\,r$, obtained from the 2-body Dirac equation with in-medium potentials~\cite{Shi:2019tji,Crater:1983ew,Crater:1987hm,Sazdjian:1986aw,Sazdjian:1988be}. We take the quark masses $m_{u,d}=0.2$ GeV, $m_s=0.3$ GeV, and $m_c=1.5$ GeV. These values are slightly smaller than the constituent quark masses due to the chiral symmetry partially restoration around the phase transition boundary, as shown in the Nambu--Jona--Lasinio model~\cite{Klevansky:1992qe}. For charmed baryons, we first combine two quarks into a diquark and then couple it with the third quark.

In the conventional simultaneous baseline, all charm-hadron species are formed on the same $\Tc$ hypersurface~\cite{Bazavov:2018mes}. In the sequential scenario, the formation temperatures follow the hierarchy obtained from the Dirac equation~\cite{Shi:2019tji},
\begin{equation}
T_{\Ds} \simeq 1.2\Tc > T_{\Dz}, T_{\Omega_c}, T_{\Xi_c}, T_{\Lambda_c} \simeq \Tc,
\end{equation}
meaning that $\Ds$ is evaluated on an earlier hypersurface while $\Dz$ freezes out at $\Tc$. The choice of $1.2\Tc$ for $\Ds$ is motivated by its larger binding energy relative to $\Dz$, which allows the $c\bar{s}$ pair to become bound earlier as the QGP cools through the crossover region. When a charm quark reaches the $1.2\Tc$ hypersurface, its coalescence probability $P_{\Ds}$ into $\Ds$ is computed from the coalescence integral. If coalescence occurs, the charm quark exits the partonic phase; otherwise, it continues to propagate until $\Tc$, where it can coalesce into the other charm-hadron species, with the coalescence probabilities rescaled by the remaining probability $1-P_{\Ds}$. Charm hadrons assigned to the earlier hypersurface are allowed to coalesce first, and the charm quarks consumed in this step are removed from the available charm-quark pool before later hadron species are constructed. In the sequential scenario, the coalescence component of $\Ds$ is therefore formed exclusively on the earlier $1.2\Tc$ hypersurface. This sequential sampling enforces charm-number conservation throughout the hadronization procedure,
\begin{eqnarray}
N_c^{\mathrm{tot}} = N_c^{\mathrm{remain}} + \sum_H N_H n_c(H),
\end{eqnarray}
where $n_c(H)$ is the number of charm quarks carried by hadron $H$. 
Charm quarks that do not coalesce on either hypersurface hadronize through fragmentation at $\Tc$, which is described by the Peterson fragmentation function ~\cite{Peterson:1982ak},
\begin{equation}
\mathscr{D}_{c\to H}(z) \propto \frac{1}{z\left(1-\frac{1}{z}-\frac{\epsilon}{1-z}\right)^2},
\end{equation}
with $z$ the momentum fraction of the charmed hadron and $\epsilon=0.01$ for mesons and $\epsilon=0.02$ for baryons~\cite{Das:2016llg}. The fragmentation fractions into various charmed hadrons are extracted from experimental data~\cite{Lisovyi:2015uqa}. The $\Ds$ mesons produced through fragmentation are also formed at $\Tc$, at the final hadronization stage.

The quantitative value $T_{\Ds} \simeq 1.2\Tc$ carries a theoretical uncertainty tied to the in-medium potential model used in the Dirac equation. If $T_{\Ds}$ were closer to $\Tc$, the hadronization-time separation $\Delta\tau(\Dz-\Ds)$ would shrink and the sequential signal in $\Delta\vtwo$ would diminish, smoothly recovering the simultaneous baseline. Conversely, a larger separation would increase $\Delta\tau$ and amplify the splitting. We varied the $\Ds$ formation temperature from $1.15T_c$ to $1.25T_c$ around the default value $1.2T_c$. The positive $v_2(\Dz)-v_2(\Ds)$ splitting in the intermediate-$p_T$ region remains qualitatively unchanged within this range.

After hadronization, the formed $D$ mesons may undergo rescattering in the hadronic phase until kinetic freeze-out at $T_{\rm kin}=137$~MeV~\cite{Pang:2018zzo}. This stage is described by a Langevin equation with a temperature-dependent spatial diffusion coefficient for $D$-meson interactions with the light-flavor hadron gas~\cite{He:2012df,He:2019tik,Torres-Rincon:2021yga}. In the present work, the hadronic diffusion of $\Dz$ is included, which further increases its elliptic flow, whereas the hadronic diffusion of $\Ds$ is neglected because of its much smaller scattering cross section~\cite{He:2012df}. As shown in our preceding work~\cite{Xu:2025ivv}, however, this effect alone is not sufficient to reverse the flow hierarchy in the simultaneous-hadronization scenario, where $\vtwo(\Dz)$ remains smaller than $\vtwo(\Ds)$.

Before presenting the results, it is instructive to clarify how the two hadronization mechanisms shape the $\Dz$--$\Ds$ flow difference. Since hadrons produced by coalescence and fragmentation carry distinctly different $\vtwo$, the species dependence of the meson $\vtwo$ receives contributions both from the different coalescence fractions of $\Dz$ and $\Ds$ and from their different formation times. To disentangle these two ingredients, we decompose the meson yields and $\vtwo$ into the two hadronization channels. The coalescence and fragmentation fractions of $\Dz$ (upper panel) and $\Ds$ (lower panel) are shown in Fig.~\ref{fig:comment1_fraction}, with the simultaneous and sequential scenarios displayed together in each panel, and the decomposition of the $\vtwo$ of $\Ds$ into the two channels, compared with the total $\vtwo$ of $\Dz$, is shown in Fig.~\ref{fig:comment1_v2} for both scenarios. In both scenarios, $\Ds$ has a larger coalescence fraction than $\Dz$, while $\Dz$ has a larger fragmentation fraction. Compared with the simultaneous baseline, the sequential scenario enhances the coalescence fraction of $\Ds$, since its coalescence probability is evaluated on the earlier and hotter $1.2\Tc$ hypersurface, and reduces that of $\Dz$ at intermediate $\pt$, because the charm quarks consumed by the early $\Ds$ formation are removed from the charm-quark pool before $\Dz$ hadronizes at $\Tc$.
\begin{center}
    \includegraphics[width=0.95\linewidth]{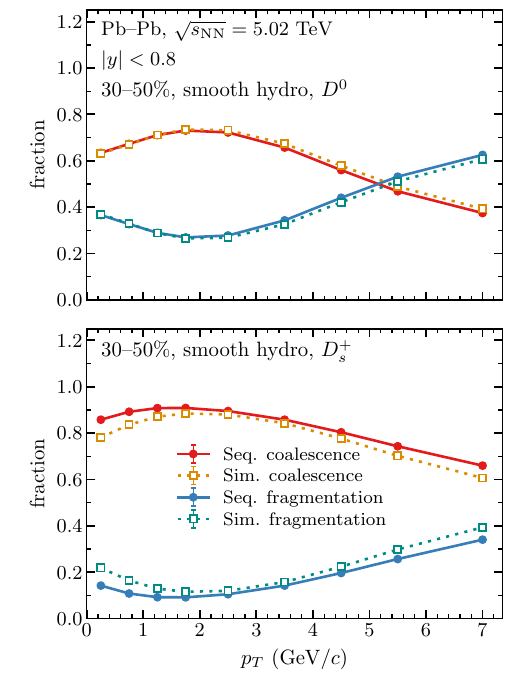}

    \captionof{figure}{
    Coalescence and fragmentation fractions of $\Dz$ (upper panel) and $\Ds$ (lower panel) as functions of $\pt$ in the 30--50\% centrality class, with the simultaneous and sequential scenarios shown together in each panel.
    }
    \label{fig:comment1_fraction}
\end{center}

In the simultaneous scenario, the coalescence component of $\Ds$ carries a much larger $\vtwo$ than its fragmentation component, and the total $\vtwo$ of $\Dz$, which receives a larger fragmentation contribution, lies below that of $\Ds$, giving $\vtwo(\Dz)<\vtwo(\Ds)$. In the sequential scenario, although the coalescence fraction of $\Ds$ is enhanced, its coalescence component carries a visibly smaller $\vtwo$ at intermediate $\pt$ than in the simultaneous case, because $\Ds$ forms earlier at $1.2\Tc$; as a result, the total $\vtwo$ of $\Ds$ drops below that of $\Dz$ at intermediate $\pt$, giving $\vtwo(\Dz)>\vtwo(\Ds)$. Note that the changes in the coalescence fractions alone would act in the opposite direction: the enhanced coalescence fraction of $\Ds$ would increase its $\vtwo$, and the reduced coalescence fraction of $\Dz$ would decrease $\vtwo(\Dz)$. The hierarchy reversal therefore demonstrates that the different coalescence/fragmentation fractions affect the magnitude of the final $\vtwo$, while the reduced partonic $\vtwo$ accumulation caused by the earlier formation of $\Ds$ is the dominant origin of the flow difference between $\Dz$ and $\Ds$ in the sequential scenario.

\begin{center}
    \includegraphics[width=0.95\linewidth]{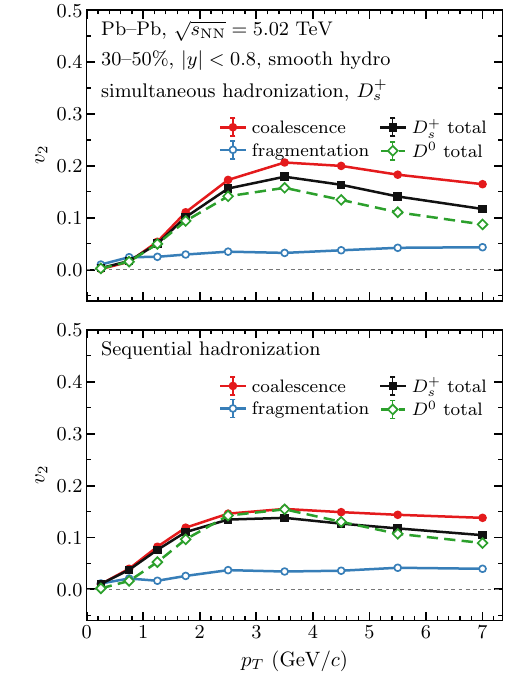}

    \captionof{figure}{
    Coalescence and fragmentation components and the total $\vtwo$ of $\Ds$, compared with the total $\vtwo$ of $\Dz$ (green dashed line), for simultaneous (upper panel) and sequential (lower panel) hadronization in the 30--50\% centrality class.
    }
    \label{fig:comment1_v2}
\end{center}

The reduced partonic accumulation identified above can be seen directly at the parton level. The charm-quark $\vtwo$ continues to develop during its propagation in the QGP and is larger at $\Tc$ than at $1.2\Tc$, particularly at low and intermediate $\pt$. In comparison, the $u,d$-quark $\vtwo$ at $\Tc$ and the strange-quark $\vtwo$ at $1.2\Tc$ are close to each other. Thus, the main partonic-level difference comes from the additional elliptic flow accumulated by charm quarks during the evolution from $1.2\Tc$ to $\Tc$ ~\cite{Xu:2025ivv}.

\section{Results and Discussion}
\label{sec:results}

Experimental measurements by ALICE have already established that the event-shape selection in the soft sector is transmitted to open-charm mesons, exhibiting a pronounced $\qtwo$ dependence of the charm-meson elliptic flow in Pb--Pb collisions at $\sqrt{s_{\mathrm{NN}}}=5.02$~TeV~\cite{ALICE:2018gif,ALICE:2020iug}. Previous theoretical work in the Catania transport framework~\cite{Sambataro:2022sns} has examined the ESE response of $D$ mesons within the conventional simultaneous-hadronization picture, showing good agreement with ALICE data and supporting the strong coupling between charm quarks and the QGP. The present study goes further by introducing sequential hadronization and demonstrating that the resulting species-dependent $\qtwo$ response---particularly the $\Dz$-$\Ds$ splitting---provides a novel handle to discriminate the space-time structure of charm freeze-out.

\subsection{Flow splitting}
\label{sec:flow-splitting}

We first examine whether the soft-sector event-shape selection is transferred to open-charm mesons. Fig.~\ref{fig:dmeson-v2} shows the $\pt$-differential elliptic flow of $\Dz$ and $\Ds$ mesons in the 0--10\% and 30--50\% centrality classes. We use 0.5~GeV/$c$ bins below $\pt=2$~GeV/$c$, and $2$--$3$, $3$--$4$, $4$--$6$, and $6$--$8$~GeV/$c$ bins above $2$~GeV/$c$, following the ALICE experimental binning so that the results can be compared directly with the data. For both species, the large-$\qtwo$ curves are well separated from the small-$\qtwo$ curves, and the separation is much larger in the 30--50\% class. The calculated $\Dz$ results also follow the ordering and magnitude of the available ALICE large- and small-$\qtwo$ data~\cite{ALICE:2018gif,ALICE:2020iug} and are consistent with the Catania transport calculation~\cite{Sambataro:2022sns}, supporting the reliability of the baseline charm-transport calculation. On this basis, the corresponding $\Ds$ curves provide a prediction for the earlier-produced strange open-charm sector.

The relative ordering of $\Dz$ and $\Ds$ depends on both $\pt$ and the event-shape class. In the sequential scenario, at very low $\pt$, $\Ds$ has a larger $\vtwo$ than $\Dz$, while at intermediate $\pt$ (coalescence dominant region) the $\Dz$ curve rises above the $\Ds$ curve, especially in the large-$\qtwo$ selection. In the 0--10\% panel (upper), this species separation is clear for large-$\qtwo$ events but much weaker for small-$\qtwo$ events. In the 30--50\% panel (lower), the separation is present for both event-shape classes, with the large-$\qtwo$ splitting still dominant; even the small-$\qtwo$ class, however, exhibits a visible $\Dz-\Ds$ separation that is more pronounced than in the 0--10\% small-$\qtwo$ case. This raises the central question: why does the same small-$\qtwo$ selection expose a stronger species dependence in semi-central collisions than in the 0--10\% class, and why does the earlier-produced strange charm meson respond differently from the non-strange one? This pattern suggests that ESE is not simply a common rescaling of all open-charm mesons. It motivates extracting the elliptic-flow difference $\Delta\vtwo(\Dz-\Ds)$ as a more direct observable for the sequential-hadronization signal.

\begin{center}
    \centering
    \includegraphics[width=0.95\linewidth]{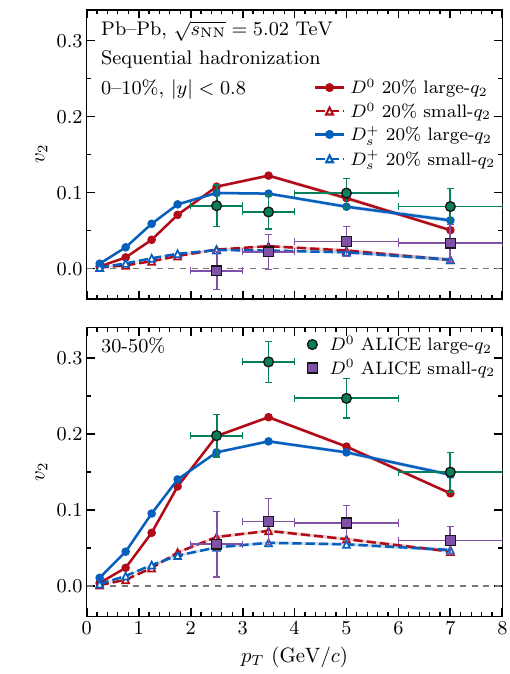}
    \captionof{figure}{The $\vtwo(\pt)$ of $\Dz$ and $\Ds$ for sequential hadronization in the 0--10\% (upper panel) and 30--50\% (lower panel) centrality classes, comparing the 20\% large-$\qtwo$ (solid lines) and small-$\qtwo$ (dashed lines) event classes.The experimental data are from the ALICE Collaboration~\cite{ALICE:2018gif,ALICE:2020iug}. The shaded bands indicate the statistical uncertainties from the finite event sample.
}
    \label{fig:dmeson-v2}
\end{center}

For comparison, we also show in Fig.~\ref{fig:comment1_simultaneous} the corresponding $\vtwo(\pt)$ results for simultaneous hadronization, where $\Dz$ and $\Ds$ hadronize on the same $\Tc$ hypersurface, with the same large-$\qtwo$ and small-$\qtwo$ event selections applied in both centrality classes. In this case, the $\vtwo$ of $\Ds$ remains larger than that of $\Dz$, and no hierarchy reversal occurs in any of the event classes.

\begin{center}
\includegraphics[width=0.95\linewidth]{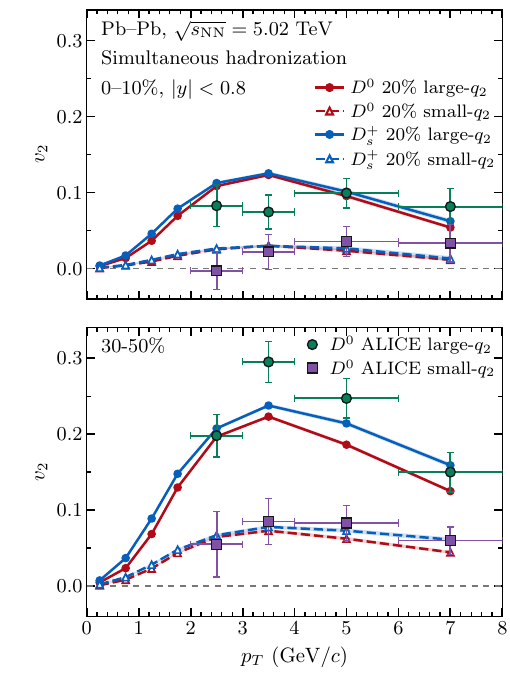}

\captionof{figure}{
The $\vtwo(\pt)$ of $\Dz$ and $\Ds$ for simultaneous hadronization in the 0--10\% (upper panel) and 30--50\% (lower panel) centrality classes, comparing the 20\% large-$\qtwo$ (solid lines) and small-$\qtwo$ (dashed lines) event classes. The ALICE data for $\Dz$ in the corresponding event classes are shown for comparison.
}
\label{fig:comment1_simultaneous}
\end{center}

Guided by the species-dependent pattern in Fig.~\ref{fig:dmeson-v2}, we next construct the flow difference $\Delta\vtwo(\Dz-\Ds)$ as the observable that directly isolates the separation between the non-strange and the earlier-produced strange charm mesons. As shown in the upper panel of Fig.~\ref{fig:delta-v2}, the sequential scenario gives a positive $\vtwo(\Dz)-\vtwo(\Ds)$ mainly in the intermediate-$\pt$ region, approximately $2.5<\pt<5$~GeV/$c$. At lower $\pt$, the $\vtwo$ of both species is dominated by their coalescence components. In this region, the sequential scenario slightly enhances the $\vtwo$ carried by the coalescence component of $\Ds$, which, together with its larger coalescence fraction, leads to $\vtwo(\Ds)>\vtwo(\Dz)$, following and even deepening the hierarchy of the simultaneous baseline.Toward intermediate $\pt$, the reduced partonic $\vtwo$ accumulation of the earlier-formed $\Ds$ overtakes this fraction-driven hierarchy and reverses the sign of the splitting. At higher $\pt$ ($\pt > 10$~GeV/$c$), the splitting decreases and turns slightly negative as well, approaching the same baseline hierarchy. This is because the fragmentation component --- for which both species are produced on the same $\Tc$ hypersurface, with no formation-time difference --- becomes increasingly important toward high $\pt$, while for the remaining coalescence component the additional partonic $\vtwo$ accumulated between $1.2\Tc$ and $\Tc$ is small at high $\pt$. The formation-time difference is therefore reflected most clearly in the intermediate-$\pt$ region.

The splitting is larger in large-$\qtwo$ events than in small-$\qtwo$ events and reaches its maximum around $\pt\simeq3$--4~GeV/$c$, reflecting the larger elliptic eccentricity selected by the event-shape procedure. This indicates in a more transparent way what is only apparent when comparing the two species in Fig.~\ref{fig:dmeson-v2}: ESE enhances the $\Dz-\Ds$ flow separation, and the enhancement is strongest in the semi-central geometry.

\begin{center}
    \centering
    \includegraphics[width=0.95\linewidth]{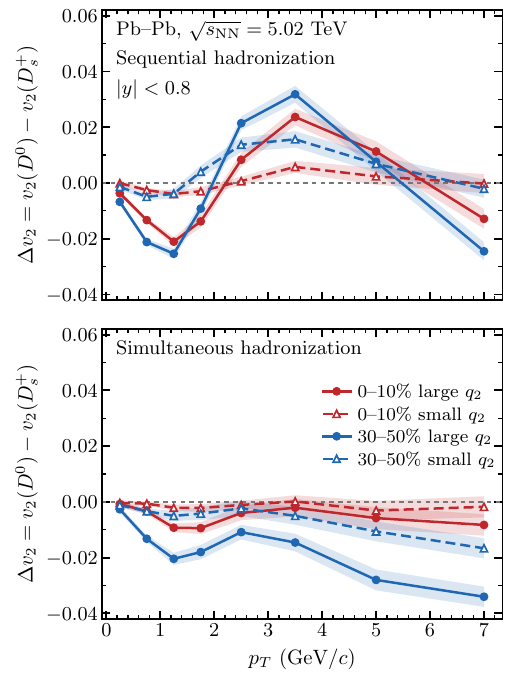}
    \captionof{figure}{$\vtwo(\Dz)-\vtwo(\Ds)$ obtained with the refined $\pt$ binning, under sequential (upper panel) and simultaneous (lower panel) hadronization, with the 0--10\% (red) and 30--50\% (blue) centrality classes overlaid in each panel; solid lines with filled circles and dashed lines with open triangles denote the large-$\qtwo$ and small-$\qtwo$ selections, respectively. The shaded bands indicate the statistical uncertainties from the finite event sample.
}
    \label{fig:delta-v2}
\end{center}

The lower panel provides the essential control test. In the simultaneous-hadronization baseline, the $\Delta\vtwo(\Dz-\Ds)$ values for the four scenarios are close to zero or negative, with the 30--50\% large-$\qtwo$ class being the most negative and the 0--10\% small-$\qtwo$ class closest to zero. They exhibit none of the positive intermediate-$\pt$ enhancement found in the sequential calculation. Therefore, the positive splitting in the upper panel is not a trivial consequence of the centrality or $\qtwo$ selection. Rather, it appears only when $\Ds$ and $\Dz$ leave the partonic phase at different times. This makes $\Delta\vtwo(\Dz-\Ds)$ an observable sensitive to sequential hadronization within the present framework. We have also verified that the normalized difference $[\vtwo(\Dz)-\vtwo(\Ds)]/[\vtwo(\Dz)+\vtwo(\Ds)]$ leads to the same qualitative conclusion: it is positive in the intermediate-$\pt$ region in the sequential scenario, while it remains consistent with zero or negative in the simultaneous scenario. Even if some simultaneous-hadronization model could reproduce a small positive $\Delta\vtwo$ at a single $\qtwo$ by adjusting transport or coalescence parameters, it would still lack the finite formation-time interval required to generate the characteristic $\qtwo$-dependence seen in the sequential scenario (see Fig.~\ref{fig:q2-v2}).

The $\qtwo$ dependence is examined in the $3.25<\pt<3.75$~GeV/$c$ interval, where the positive sequential $\Delta\vtwo(\Dz-\Ds)$ is already prominent and rising toward its maximum (see Fig.~\ref{fig:delta-v2}). We examine in Fig.~\ref{fig:q2-v2} whether this separation evolves systematically with the event-shape variable, rather than only between two coarse large- and small-$\qtwo$ classes. In the upper panel, $\vtwo$ increases with $\qtwo$ for both $\Dz$ and $\Ds$. The 30--50\% points occupy both larger $\qtwo$ and larger $\vtwo$ values than the 0--10\% points, consistent with the stronger elliptic geometry in semi-central collisions. Within a given centrality class, the $\Dz$ points rise more strongly with $\qtwo$ than the $\Ds$ points, which is the differential response anticipated from Fig.~\ref{fig:dmeson-v2}.
\begin{center}
    \centering
    \includegraphics[width=0.95\linewidth]{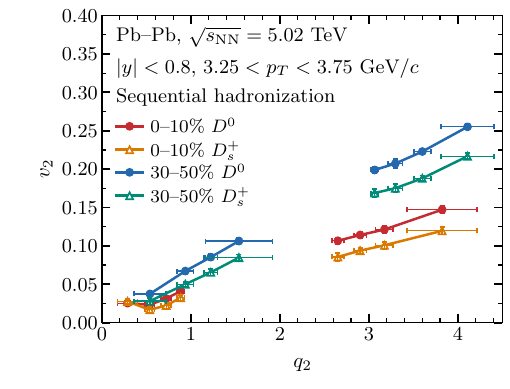}
    \vspace{0.4em}
    \includegraphics[width=0.95\linewidth]{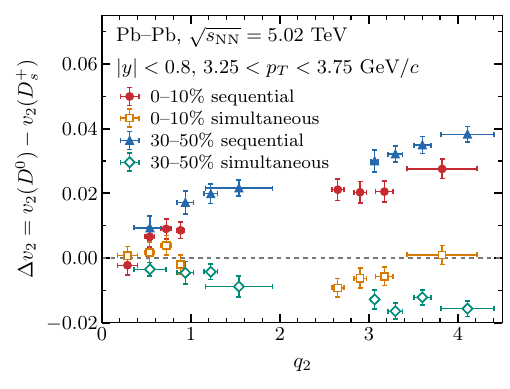}
    \captionof{figure}{$\qtwo$ dependence in $3.25<\pt<3.75$ GeV/$c$. Upper panel: $\vtwo\{\mathrm{SP}\}$ of $\Dz$ and $\Ds$ in the sequential scenario. Lower panel: $\Delta\vtwo(\Dz-\Ds)$ compared between sequential and simultaneous hadronization.}
    \label{fig:q2-v2}
\end{center}
The lower panel of Fig.~\ref{fig:q2-v2} translates this behavior into $\Delta\vtwo(\Dz-\Ds)$. In the sequential calculation, the splitting is positive and increases with $\qtwo$ in both centrality classes, with 30--50\% systematically above 0--10\%. In the simultaneous calculation, the 0--10\% points stay close to zero, while the 30--50\% points are negative; neither exhibits a pronounced $\qtwo$ dependence comparable to the sequential scenario. This contrast answers the question raised by Fig.~\ref{fig:dmeson-v2}: the stronger semi-central species separation is not simply caused by choosing small or large $\qtwo$ events, but by how the ESE-selected geometry is converted into a species-dependent flow response when the two mesons hadronize sequentially. The opposite $\qtwo$ trends in the two scenarios demonstrate that ESE does not merely rescale an existing splitting; it exposes the distinct functional dependence introduced by sequential freeze-out.

The next question is whether the stronger $\Delta\vtwo(\Dz-\Ds)$ in semi-central collisions simply comes from a larger hadronization-time gap. Fig.~\ref{fig:formation-time} shows that this is not the case. In the sequential calculation, $\Delta\tau(\Dz-\Ds)$ is positive over the full momentum range, supporting that $\Dz$ forms later than $\Ds$. The time gap is largest at low $\pt$ and decreases monotonically with increasing $\pt$. It is also larger in 0--10\% collisions than in 30--50\% collisions, while the large- and small-$\qtwo$ selections within the same centrality class almost overlap. The near overlap of the large-$\qtwo$ and small-$\qtwo$ curves within the same centrality class shows that the ESE selection does not produce a noticeably different hadronization-time interval or extend the interaction duration. The clear $\qtwo$ dependence observed in the flow observables therefore mainly reflects the change in the initial geometry and the corresponding in-medium response. The simultaneous baseline stays slightly below zero, as expected when the two species hadronize on the same hypersurface. Here, $\Delta\tau$ is defined from the mean formation times of the finally produced $\Dz$ and $\Ds$ mesons. A small nonzero value can therefore arise from species-dependent sampling even when both species hadronize at the same $\Tc$; its magnitude, $|\Delta\tau|\lesssim0.2$~fm/$c$, is much smaller than the sequential result.

This comparison separates the two ingredients of the signal. The sequential scenario supplies a finite $\Ds$-to-$\Dz$ time ordering, but the ESE selection does not by itself generate a different formation time. Moreover, the central 0--10\% class has the larger $\Delta\tau(\Dz-\Ds)$ but not the larger flow splitting. Therefore the observed hierarchy of $\Delta\vtwo(\Dz-\Ds)$ cannot be read as a direct map of the time gap alone. It depends on how much elliptic flow can still be accumulated during that interval, which is controlled by the geometry and response of the medium. Within the present framework, the weak $\qtwo$ dependence of $\Delta\tau$ together with the clear $\qtwo$ dependence of the flow observables indicates that ESE primarily probes the geometry-dependent medium response rather than a change in the hadronization-time interval.

\begin{center}
    \centering
    \includegraphics[width=1.0\linewidth]{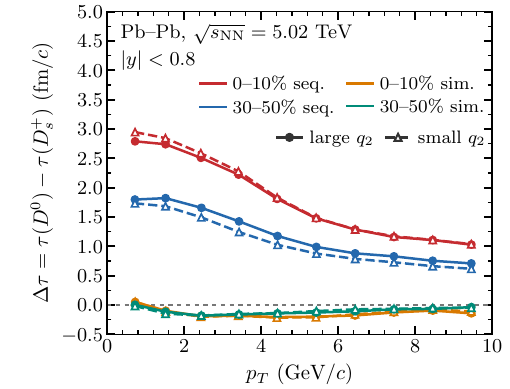}
    \captionof{figure}{Hadronization-time difference $\Delta\tau(\Dz-\Ds)$ as a function of $\pt$. Sequential hadronization gives a finite positive time gap, while the simultaneous baseline remains near zero.}
    \label{fig:formation-time}
\end{center}

To quantify this conversion from event geometry to final charm flow, Fig.~\ref{fig:ese-response} shows the $\qtwo$-response slope $\chi$. In each ($\pt$, centrality) bin, the eight equal-size $\qtwo$ subsamples are constructed within the selected 20\% large-$\qtwo$ and 20\% small-$\qtwo$ event classes. Events in the intermediate 60\% of the $\qtwo$ distribution are excluded from the fit; the mean $\vtwo$ in each sub-sample is then fitted to the linear form $\vtwo\{\mathrm{SP}\} = \chi\,\qtwo + v_2^{(0)}$ over the selected $\qtwo$ range, where $v_2^{(0)}$ is the intercept. This selection follows the experimental ESE strategy and maximizes the event-shape contrast between the two classes. The fitted slopes, their uncertainties, the intercepts, the $R^2$ values, and the fit residuals are summarized in Tables~\ref{tab:v2_q2_fit_010_sequential} and~\ref{tab:v2_q2_fit_3050_sequential}, and the corresponding fit curves are shown in Figs.~\ref{fig:comment5_v2q2_fit} and~\ref{fig:comment5_v2q2_fit_3050}. 
The approximately linear dependence holds in the intermediate-$\pt$ region where the sequential-hadronization signal is studied, with $R^2\gtrsim0.98$ for both species in both centrality classes (see Tables~\ref{tab:v2_q2_fit_010_sequential} and~\ref{tab:v2_q2_fit_3050_sequential}); at low $\pt$, the larger statistical fluctuations reduce the fit quality. Unlike $\Delta\vtwo$, which depends on the absolute magnitude of the event geometry, $\chi$ isolates the susceptibility of each species to geometric changes---a quantity that is meaningful only under active ESE selection. Importantly, the Compact Muon Solenoid (CMS) Collaboration has recently measured $\Dz$-meson $\vtwo$ in different $\qtwo$ event classes and further studied its correlation with the charged-particle $\vtwo$~\cite{CMS:2026ese}; the slope $\chi$ can be extracted directly from existing data without requiring any model-dependent reconstruction of the hadronization time.

The upper panel of Fig.~\ref{fig:ese-response} shows
the hierarchy $\chi(\Dz)>\chi(\Ds)$ in the
intermediate-$\pt$ region of the sequential scenario. The difference between $\chi(\Dz)$ and $\chi(\Ds)$ is
largest around $\pt\simeq3$--$3.5$~GeV/$c$, the same
momentum region where the positive
$\Delta\vtwo(\Dz-\Ds)$ is largest, while the baseline
ordering $\chi(\Ds)\geq\chi(\Dz)$ is restored toward
both lower and higher $\pt$. This is a quantitative, and falsifiable prediction: if future analyses of the ALICE data were to find $\chi(\Dz) \approx \chi(\Ds)$, the sequential picture would be disfavored. The simultaneous baseline makes the opposite prediction. In the 0--10\% class, the simultaneous $\Delta\vtwo$ shows almost no $\qtwo$ dependence (lower panel of Fig.~\ref{fig:q2-v2}), implying $\chi(\Dz) - \chi(\Ds) \approx 0$; in the 30--50\% class, the simultaneous $\Delta\vtwo$ becomes more negative toward larger $\qtwo$ (lower panel of Fig.~\ref{fig:q2-v2}), yielding $\chi(\Dz) - \chi(\Ds) < 0$. Thus, in the intermediate-$\pt$ region, the sign of
$\chi(\Dz)-\chi(\Ds)$ distinguishes the two hadronization
scenarios within the present calculation: the sequential
scenario gives a positive difference, whereas the
simultaneous baseline gives a difference close to zero
or negative.

\begin{center}
    \centering
    \includegraphics[width=1.0\linewidth]{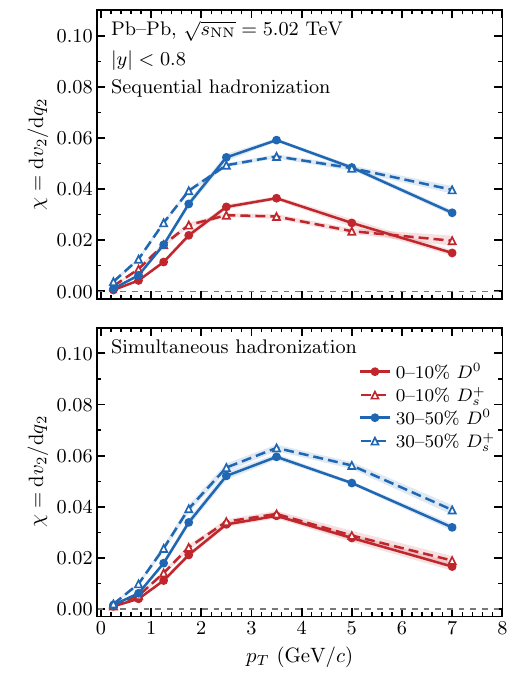}

\captionof{figure}{The ESE response coefficient $\chi(\pt)$ of $\Dz$ (solid lines with filled circles) and $\Ds$ (dashed lines with open triangles) over the extended $\pt$ range, for sequential (upper panel) and simultaneous (lower panel) hadronization in the 0--10\% (red) and 30--50\% (blue) centrality classes. The shaded bands indicate the uncertainties of the fitted slopes.}
    \label{fig:ese-response}
\end{center}

All extracted slopes are positive, so both $\Dz$ and $\Ds$ follow the ESE-selected geometry. The response in 30--50\% collisions is much larger than in 0--10\% collisions, reflecting a higher conversion efficiency from initial eccentricity to final charm flow: the 30--50\% geometry provides both a stronger lever arm and a QGP lifetime that remains long enough for charm quarks to fully respond, whereas the smaller initial eccentricity in central collisions suppresses the conversion even though the system lives longer. Because $\Dz$ forms later, it samples this late-stage response more strongly than $\Ds$. This response hierarchy answers the puzzle raised by Fig.~\ref{fig:dmeson-v2}. Even for a small-$\qtwo$ selection, the semi-central medium converts the remaining elliptic geometry---still present among low-$\qtwo$ events---into charm-meson flow more efficiently than the central medium. The visible $\Dz-\Ds$ separation in 30--50\% small-$\qtwo$ events is therefore not contradictory to the small-$\qtwo$ label; it reflects a larger response coefficient in a more eccentric collision geometry.

\begin{widetext}

\begin{center}
\includegraphics[width=0.85\textwidth]{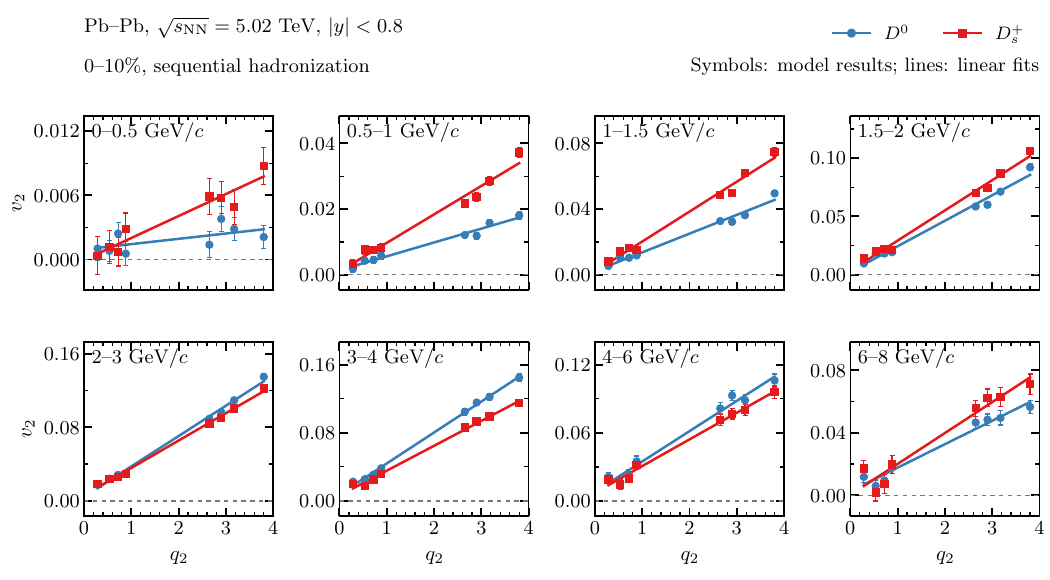}
\captionof{figure}{The $\vtwo$--$\qtwo$ relations and the corresponding linear fits in the refined $\pt$ bins for $\Dz$ (blue) and $\Ds$ (red) in 0--10\% Pb--Pb collisions with sequential hadronization. The points show the mean $\vtwo$ in the eight $\qtwo$ subsamples, and the lines show the linear fits $\vtwo=\chi\,\qtwo+v_2^{(0)}$.}
\label{fig:comment5_v2q2_fit}
\end{center}

\begin{center}
\captionof{table}{Linear-fit parameters for $v_2(q_2)=\chi q_2+v_2^0$ in 0--10\% Pb--Pb collisions with sequential hadronization. For each species and $\pt$ interval, the table lists the slope $\chi$, its uncertainty $\sigma_\chi$, the intercept $v_2^0$, the coefficient of determination $R^2$, and the residuals $r_i=v_2^i-v_2^{\rm fit}$ of the eight $q_2$ subsamples. The values of $\chi$, $\sigma_\chi$, $v_2^0$, and $r_i$ are given in units of $10^{-3}$.}
\label{tab:v2_q2_fit_010_sequential}

\scriptsize
\setlength{\tabcolsep}{2.2pt}
\renewcommand{\arraystretch}{1.0}
\resizebox{\textwidth}{!}{%
\begin{tabular}{ccrrrrrrrrrrrr}
\hline

Species & $p_T$ (GeV/$c$) & $\chi$ & $\sigma_{\chi}$ & $v_2^0$ & $R^2$

& $r_1$ & $r_2$ & $r_3$ & $r_4$ & $r_5$ & $r_6$ & $r_7$ & $r_8$ \\

\hline

$D^0$   & 0--0.5 & 0.495 & 0.260 & 0.946 & 0.3774 & -0.065 & -0.397 &  1.126 & -0.832 & -0.873 &  1.418 &  0.345 & -0.722 \\

$D^0$   & 0.5--1 & 4.220 & 0.273 & 1.376 & 0.9756 & -0.784 &  0.622 &  0.098 &  0.662 & -0.473 & -1.748 &  0.971 &  0.652 \\

$D^0$   & 1--1.5 & 11.429 & 0.648 & 2.177 & 0.9811 & -0.124 &  1.917 & -0.076 & -0.560 &  0.202 & -3.106 & -2.227 &  3.974 \\

$D^0$   & 1.5--2 & 21.865 & 1.131 & 2.534 & 0.9842 &  0.858 &  3.933 & -0.145 & -2.513 & -1.970 & -5.978 & -0.694 &  6.509 \\

$D^0$   & 2--3   & 32.994 & 0.966 & 4.483 & 0.9949 &  3.368 &  1.908 & -0.051 & -3.965 & -2.717 & -3.675 &  0.142 &  4.990 \\

$D^0$   & 3--4   & 36.398 & 0.718 & 7.563 & 0.9977 &  4.536 & -1.620 & -2.976 & -1.421 &  0.798 &  2.449 & -0.733 & -1.031 \\

$D^0$   & 4--6   & 26.725 & 1.549 & 7.994 & 0.9803 &  5.531 & -6.701 & -4.539 &  3.160 &  2.723 &  7.383 & -4.117 & -3.440 \\

$D^0$   & 6--8   & 14.984 & 1.085 & 2.719 & 0.9695 &  4.804 & -4.728 & -4.040 &  1.656 &  4.114 &  2.148 & -0.803 & -3.151 \\

\hline

$D_s^+$ & 0--0.5 & 2.053 & 0.255 & -0.040 & 0.9150 & -0.187 &  0.100 & -0.729 &  1.071 &  0.484 & -0.140 & -1.564 &  0.965 \\

$D_s^+$ & 0.5--1 & 8.669 & 0.564 & 1.021 & 0.9752 & -0.144 &  1.962 &  0.225 & -0.616 & -2.289 & -2.464 &  0.085 &  3.241 \\

$D_s^+$ & 1--1.5 & 18.237 & 0.886 & 1.956 & 0.9860 &  0.786 &  2.465 &  1.207 & -3.129 & -1.708 & -5.178 &  1.959 &  3.599 \\

$D_s^+$ & 1.5--2 & 25.889 & 1.012 & 3.360 & 0.9909 &  3.198 &  2.622 &  0.161 & -5.099 & -2.144 & -4.180 &  1.292 &  4.150 \\

$D_s^+$ & 2--3   & 29.768 & 0.741 & 6.055 & 0.9963 &  3.613 &  1.613 & -1.708 & -3.293 & -0.719 & -2.308 & -0.307 &  3.110 \\

$D_s^+$ & 3--4   & 29.236 & 0.928 & 6.283 & 0.9940 &  5.445 & -4.407 & -2.256 & -0.819 &  2.411 &  2.094 & -0.106 & -2.361 \\

$D_s^+$ & 4--6   & 23.502 & 1.068 & 7.201 & 0.9878 &  4.993 & -5.352 & -4.268 &  3.605 &  2.064 &  1.270 & -1.545 & -0.767 \\

$D_s^+$ & 6--8   & 19.759 & 1.974 & 0.310 & 0.9435 & 11.218 & -9.183 & -7.444 &  1.985 &  3.003 &  4.810 & -0.149 & -4.239 \\

\hline
\end{tabular}%
}
\end{center}

\begin{center}
\includegraphics[width=0.85\textwidth]{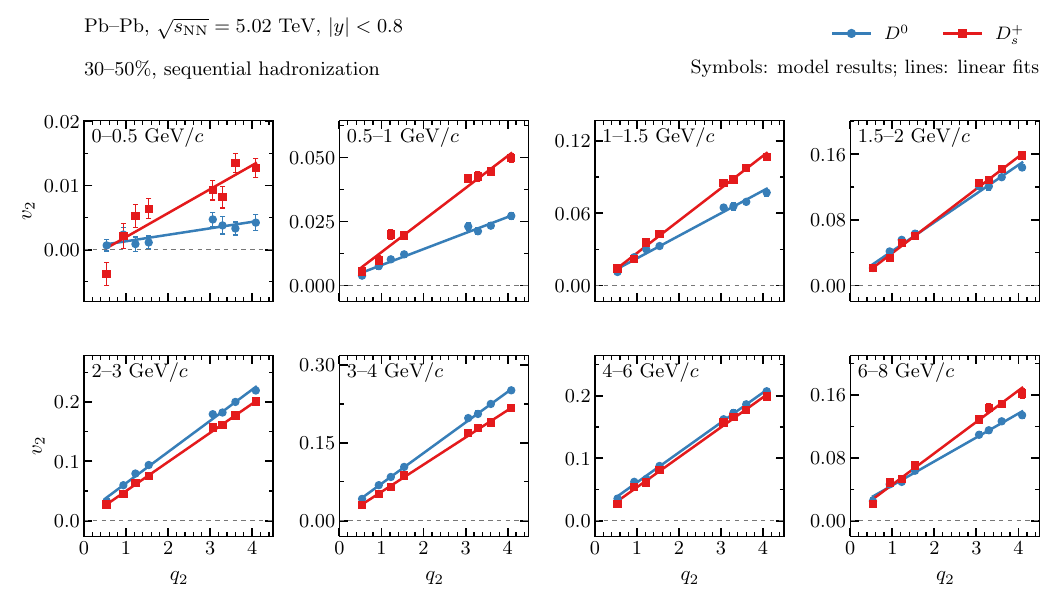}
\captionof{figure}{The $\vtwo$--$\qtwo$ relations and the corresponding linear fits in the refined $\pt$ bins for $\Dz$ (blue) and $\Ds$ (red) in 30--50\% Pb--Pb collisions with sequential hadronization. The points show the mean $\vtwo$ in the eight $\qtwo$ subsamples, and the lines show the linear fits $\vtwo=\chi\,\qtwo+v_2^{(0)}$.}
\label{fig:comment5_v2q2_fit_3050}
\end{center}

\begin{center}
\captionof{table}{Linear-fit parameters for $v_2(q_2)=\chi q_2+v_2^0$ in 30--50\% Pb--Pb collisions with sequential hadronization. For each species and $\pt$ interval, the table lists the slope $\chi$, its uncertainty $\sigma_\chi$, the intercept $v_2^0$, the coefficient of determination $R^2$, and the residuals $r_i=v_2^i-v_2^{\rm fit}$ of the eight $q_2$ subsamples. The values of $\chi$, $\sigma_\chi$, $v_2^0$, and $r_i$ are given in units of $10^{-3}$.}
\label{tab:v2_q2_fit_3050_sequential}

\scriptsize
\setlength{\tabcolsep}{2.2pt}
\renewcommand{\arraystretch}{1.0}
\resizebox{\textwidth}{!}{%
\begin{tabular}{ccrrrrrrrrrrrr}
\hline

Species & $p_T$ (GeV/$c$) & $\chi$ & $\sigma_{\chi}$ & $v_2^0$ & $R^2$

& $r_1$ & $r_2$ & $r_3$ & $r_4$ & $r_5$ & $r_6$ & $r_7$ & $r_8$ \\

\hline

$D^0$   & 0--0.5 &  1.011 & 0.237 &  0.310 & 0.7514 & -0.196 &  1.096 & -0.681 & -0.753 &  1.296 &  0.127 & -0.655 & -0.235 \\

$D^0$   & 0.5--1 &  6.243 & 0.349 &  1.824 & 0.9816 & -1.315 & -0.079 &  0.779 &  0.749 &  2.105 & -1.180 & -0.923 & -0.136 \\

$D^0$   & 1--1.5 & 18.185 & 0.752 &  5.278 & 0.9898 & -3.812 &  1.220 &  2.550 & -0.483 &  3.735 &  0.533 & -1.306 & -2.436 \\

$D^0$   & 1.5--2 & 34.152 & 1.325 &  9.445 & 0.9910 & -6.404 &  0.164 &  4.490 &  1.256 &  6.982 & -1.047 & -0.353 & -5.087 \\

$D^0$   & 2--3   & 52.374 & 1.416 & 11.185 & 0.9956 & -6.210 & -0.529 &  4.142 &  1.925 &  7.809 & -1.538 &  0.324 & -5.923 \\

$D^0$   & 3--4   & 59.138 & 0.649 & 12.068 & 0.9993 & -2.035 &  0.947 & -0.221 &  0.463 &  4.518 & -1.170 & -0.017 & -2.485 \\

$D^0$   & 4--6   & 48.409 & 0.900 & 11.817 & 0.9979 & -2.693 &  4.947 & -4.278 &  1.214 &  2.264 &  0.950 &  0.119 & -2.523 \\

$D^0$   & 6--8   & 30.693 & 1.032 & 13.439 & 0.9933 & -5.153 &  3.948 & -1.662 &  2.833 &  1.802 &  0.447 &  2.442 & -4.657 \\

\hline

$D_s^+$ & 0--0.5 &  3.767 & 0.695 & -1.915 & 0.8305 & -3.937 &  0.516 &  2.553 &  2.493 & -0.349 & -2.344 &  1.839 & -0.771 \\

$D_s^+$ & 0.5--1 & 12.562 & 0.709 &  0.486 & 0.9813 & -1.812 & -2.412 &  4.096 & -0.221 &  2.818 &  0.684 & -1.169 & -1.984 \\

$D_s^+$ & 1--1.5 & 26.776 & 0.705 &  0.308 & 0.9959 & -0.646 & -3.598 &  2.741 &  1.147 &  2.717 & -0.564 &  1.067 & -2.864 \\

$D_s^+$ & 1.5--2 & 39.320 & 0.816 &  0.248 & 0.9974 &  0.172 & -3.851 &  3.275 & -0.620 &  4.431 & -1.068 &  0.071 & -2.409 \\

$D_s^+$ & 2--3   & 49.317 & 0.761 &  0.657 & 0.9986 & -0.064 & -2.381 &  2.134 & -0.818 &  5.364 & -1.505 & -1.230 & -1.499 \\

$D_s^+$ & 3--4   & 52.762 & 0.794 &  2.936 & 0.9986 & -0.591 & -0.906 & -2.126 &  2.789 &  4.487 &  1.319 & -3.503 & -1.469 \\

$D_s^+$ & 4--6   & 48.168 & 1.057 &  5.175 & 0.9971 & -4.068 &  4.078 & -3.203 &  2.176 &  4.571 &  1.832 & -1.885 & -3.501 \\

$D_s^+$ & 6--8   & 39.689 & 1.455 &  6.190 & 0.9920 & -6.711 &  4.999 & -1.930 &  2.965 &  1.110 &  6.437 & -0.408 & -6.462 \\

\hline
\end{tabular}%
}
\end{center}

\end{widetext}

The combined set of flow observables converges on a coherent physical picture that validates the central premise of this work: event-shape engineering can separate the bulk geometry response of the medium from the hadronization-time response of individual charm-hadron species. Event-shape engineering primarily controls the initial eccentricity and hence the magnitude of the collective response, producing the expected large-$\qtwo$ enhancement of inclusive $D$-meson $\vtwo$. Sequential hadronization, however, introduces an additional knob---the time at which each species leaves the partonic phase---which modulates the relative flow of $\Dz$ and $\Ds$ independently of the global geometry. The simultaneous scenario lacks this time ordering and therefore cannot naturally generate the same positive $\Dz-\Ds$ splitting or its characteristic centrality and $\qtwo$ dependence.

The semi-central 30--50\% class emerges as the optimal window for observing this mechanism. It combines a sizeable initial eccentricity---which provides a strong lever arm for the ESE selection---with a QGP lifetime that is still long enough for the $1.2\Tc \to \Tc$ interval to generate appreciable additional charm-quark flow. In the most central class, the system may live longer, but the smaller initial eccentricity limits the conversion of late-stage interactions into additional elliptic flow, even for large-$\qtwo$ events. Conversely, in more peripheral collisions the eccentricity is larger but the medium lifetime shorter, compressing the sequential time window~\cite{Du:2026system}. This non-monotonic interplay between lifetime and eccentricity explains why the 30--50\% centrality interval offers the clearest discrimination between sequential and simultaneous hadronization, and why event-shape engineering---which actively selects on the geometry degree of freedom---is essential for exposing the effect.

The ESE analysis above does not merely amplify the sequential-hadronization signal; it provides qualitatively new information that is inaccessible in inclusive measurements. By fixing the centrality and varying only the initial eccentricity, ESE separates the bulk geometry response from the hadronization-time response, exposes a characteristic $\qtwo$ fingerprint that the simultaneous baseline cannot reproduce, and identifies the semi-central regime as the optimal window where the late-stage medium response is most efficiently converted into charm-meson flow.

\subsection{Hadron chemistry}
\label{sec:hadron-chemistry}

Before examining the event-shape dependence of charm-hadron chemistry, we first compare the $\Dz$ and $\Ds$ transverse-momentum spectra with the ALICE measurements. Figure~\ref{fig:charm_spectra_smooth} shows the total spectra and their coalescence and fragmentation components in the sequential scenario for the 0--10\% and 30--50\% centrality classes. These results are obtained with a smooth hydrodynamic background generated using the same parameter set as the event-by-event calculation.

\begin{center}
    \centering
    \includegraphics[width=1\linewidth]{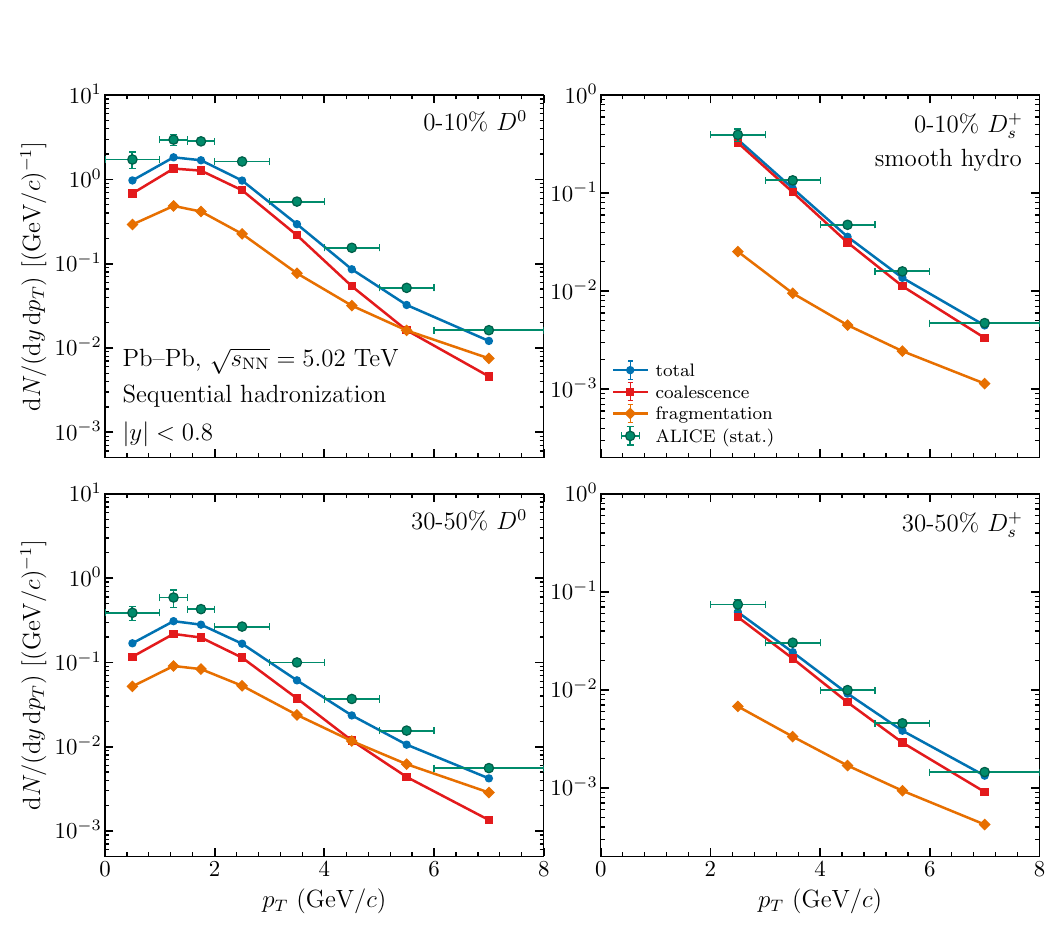}
    \captionof{figure}{Transverse-momentum spectra of $\Dz$ (left panels) and $\Ds$ (right panels) in the 0--10\% (upper panels) and 30--50\% (lower panels) centrality classes for sequential hadronization. The total, coalescence, and fragmentation contributions are calculated with the smooth hydrodynamic background, without event-shape selection, and compared with the ALICE measurements.}
    \label{fig:charm_spectra_smooth}
\end{center}

We next examine whether the ESE selection changes the relative strange-to-nonstrange charm yield. Fig.~\ref{fig:ds-over-d0} shows the total $\Ds/\Dz$
ratio and the ratios for mesons produced through
coalescence and fragmentation in the sequential scenario.
The upper and lower panels correspond to the 0--10\%
and 30--50\% centrality classes, respectively.
Within each centrality class, the large- and small-$\qtwo$
results remain close for both the total ratio and
the individual hadronization channels. Compared with the ALICE measurement~\cite{ALICE:2021rxa}, the calculation overpredicts the ratio at low $\pt$ in both centrality classes. The decomposition into the coalescence and fragmentation components shows that the low-$\pt$ enhancement originates from the coalescence component, while the fragmentation component is small and rather flat. The hadronization parameters were not retuned to reproduce the measured $\Ds/\Dz$ ratio: the relative coalescence probabilities among the different charm-hadron species and the treatment of excited-state feed-down were kept as in our existing hadronization framework. As discussed in Ref.~\cite{Xu:2025ivv}, the earlier coalescence of $\Ds$ enhances the low-$\pt$ ratio in the sequential scenario in the 0--10\% centrality class. The low-$\pt$ $\Ds/\Dz$ ratio is highly sensitive to these ingredients, which differ between model implementations~\cite{Zhao:2023nrz}; future precision measurements in this region, together with excited-state measurements, will provide strong constraints on them. We note, however, that this discrepancy does not affect the predicted flow splitting: the meson $\vtwo$ is an intensive quantity, determined by the momentum anisotropies of the coalescence and fragmentation components and by their relative weights rather than by the absolute species abundances, and, as discussed in Sec.~II , the change of the coalescence fractions in the sequential scenario acts in the direction opposite to the observed splitting.

The key point for the present ESE analysis is that the large- and small-$\qtwo$ curves within the same centrality class are close to each other. Thus, unlike $\vtwo$, the $\Ds/\Dz$ yield ratio is not strongly reorganized by the event-shape selection. We emphasize that this insensitivity is specific to the event-shape selection: comparing different centrality classes would change not only the initial eccentricity but also the hadron chemistry --- through the different fireball sizes and medium-response efficiencies --- whereas the $\qtwo$ selection within a fixed centrality class leaves the relative charm-hadron yields unchanged. This provides an important control: the large $\Dz-\Ds$ flow difference seen above is not mainly produced by a large change in the relative number of $\Ds$ and $\Dz$ mesons between large- and small-$\qtwo$ events.

\begin{center}
    \centering
    \includegraphics[width=0.95\linewidth]{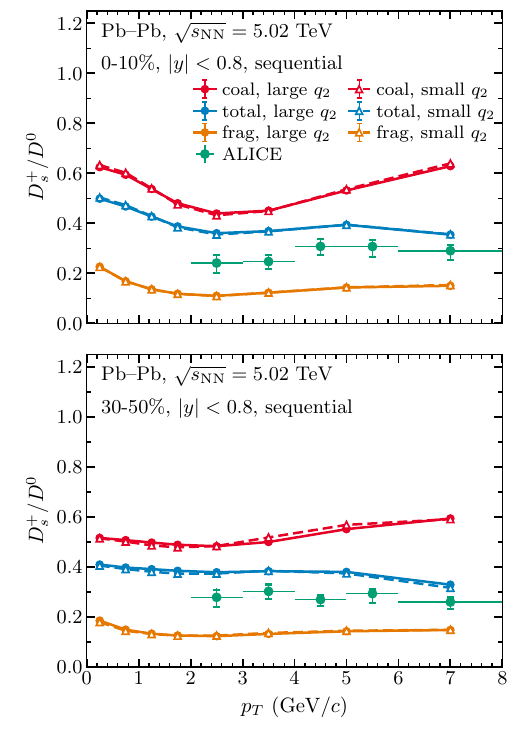}
    \captionof{figure}{$\Ds/\Dz$ ratio as a function of $\pt$ for large- and small-$\qtwo$ event classes in the 0--10\% (upper panel) and 30--50\% (lower panel) centrality classes in the sequential scenario, shown together with its coalescence and fragmentation components. The ALICE data~\cite{ALICE:2021rxa} are shown for comparison.}
    \label{fig:ds-over-d0}
\end{center}

Fig.~\ref{fig:rq2-ds-over-d0} makes this control test more explicit by forming the $\qtwo$ double ratio of $\Ds/\Dz$, defined as $R_{\qtwo} \equiv (\Ds/\Dz)_{\text{large-}\qtwo} / (\Ds/\Dz)_{\text{small-}\qtwo}$. If the event-shape selection strongly modified charm hadron chemistry, this observable would move substantially away from unity. Instead, the double ratio remains close to one over the full $\pt$ range for both centrality classes. The sequential and simultaneous calculations are also very close in this observable. Therefore, the $\qtwo$ dependence of the $\Ds/\Dz$ ratio is not a sensitive discriminator of the formation-time ordering. Yet this very insensitivity is informative: it tells us that ESE produces a much clearer signal in the anisotropic flow sector than in the integrated particle composition, and it strengthens the interpretation of $\Delta\vtwo(\Dz-\Ds)$ as a dynamical response observable.

\begin{center}
    \centering
    \includegraphics[width=1.0\linewidth]{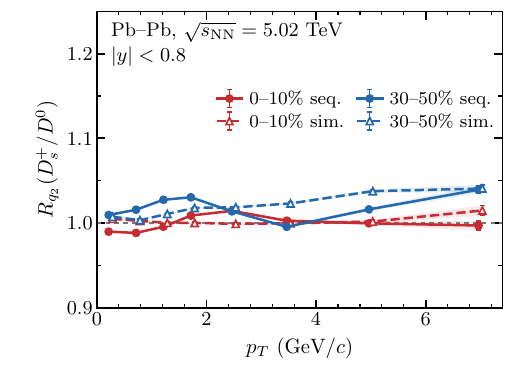}
    \captionof{figure}{$R_{\qtwo}$ of the $\Ds/\Dz$ ratio as a function of $\pt$ for sequential and simultaneous hadronization scenarios.}
    \label{fig:rq2-ds-over-d0}
\end{center}

The coalescence fraction in Fig.~\ref{fig:coalescence-fraction} explains why the strange and non-strange charm mesons are chemically different even though their $\qtwo$ yield modification is weak. In the sequential calculation, the $\Ds$ coalescence fraction is substantially larger than that of $\Dz$ over the full plotted range. Both species show a broad maximum at low-to-intermediate $\pt$ and then decrease toward high $\pt$, where fragmentation becomes increasingly important. This indicates that the $\Ds$ channel is more strongly tied to recombination with thermal strange quarks, while $\Dz$ receives a larger relative fragmentation component. At the same time, the large-$\qtwo$ and small-$\qtwo$ selections remain close for both species. The coalescence mechanism therefore sets the baseline species difference, but it does not generate a strong event-shape dependence of the $\Ds/\Dz$ chemistry. Combined with Figs.~\ref{fig:ds-over-d0} and \ref{fig:rq2-ds-over-d0}, this shows that the sequential signal has two layers: chemistry distinguishes the hadronization channels of $\Ds$ and $\Dz$, while ESE exposes how their different formation times are converted into a measurable flow splitting. This within-system control---fixing the centrality while varying only the event shape---is inaccessible in inclusive measurements and provides a stringent model-internal consistency test that the observed splitting is a dynamical flow effect rather than a chemical yield modification.

\begin{center}
    \centering
    \includegraphics[width=1.0\linewidth]{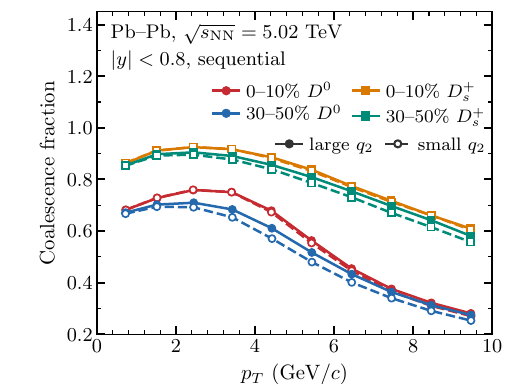}
    \captionof{figure}{Coalescence fractions of $\Dz$ and $\Ds$ as functions of $\pt$ in the sequential hadronization scenario.}
    \label{fig:coalescence-fraction}
\end{center}

\section{Summary}
\label{sec:summary}

We have presented an event-shape-engineering analysis of sequential hadronization for $\Dz$ and $\Ds$ mesons in Pb--Pb collisions at $\sqrt{s_{\mathrm{NN}}}=5.02$ TeV. The study compares large- and small-$\qtwo$ event classes in 0--10\% and 30--50\% centrality intervals and contrasts sequential and simultaneous hadronization scenarios.

The main conclusion is that the $\Dz-\Ds$ elliptic-flow splitting is controlled by a late-stage partonic evolution window between $1.2\Tc$ and $\Tc$. The calculated $\Dz$ results are consistent with available ALICE measurements, while the corresponding $\Ds$ observables constitute predictions of the sequential-hadronization framework. In the sequential scenario, $\Ds$ forms earlier and carries less late-stage charm flow, while $\Dz$ forms later after its parent charm quark has accumulated additional anisotropy. This mechanism yields a positive $\Delta\vtwo(\Dz-\Ds)$ at intermediate $\pt$ (approximately $2.5$--$5$~GeV/$c$) and a characteristic event-shape dependence, with the semi-central 30--50\% class providing the clearest discrimination. The $\qtwo$-response slope $\chi$ emerges as a particularly robust and experimentally accessible discriminator: the species-dependent hierarchy $\chi(\Dz) > \chi(\Ds)$ in the sequential scenario, together with the predicted opposite sign or near-zero difference in the simultaneous baseline, provides a direct and falsifiable test of the space-time structure of charm freeze-out. The simultaneous scenario lacks this time separation and therefore provides a natural baseline for identifying the sequential effect.

The accompanying hadronization observables support the same interpretation. The $\Ds/\Dz$ ratio reflects the enhanced strange-quark coalescence on the earlier hypersurface, and its insensitivity to the $\qtwo$ selection indicates that the splitting is a dynamical rather than chemical effect. The coalescence fraction identifies the momentum region where the mechanism is most active, and the hadronization-time difference demonstrates that the late $1.2\Tc\rightarrow\Tc$ interval remains appreciable over the relevant $\pt$ range. Beyond quantifying the sequential signal, ESE provides a qualitative discriminator: the distinct $\qtwo$ dependence of $\Delta\vtwo(\Dz-\Ds)$ in the sequential scenario exposes the functional form expected from a finite freeze-out interval, whereas the simultaneous baseline lacks this geometry-time coupling. Together, these results suggest the $\Dz-\Ds$ flow splitting under event-shape engineering as a sensitive probe of sequential hadronization and of the space-time structure of the QGP near the transition temperature.

\section*{Acknowledgments}
B.-W. Z. is supported by the National Natural Science Foundation of China with Project No. 12535010, and J. Z. is supported by the Helmholtz Research Academy Hesse for FAIR.

\section*{Data Availability}
There are no publicly available research data or software supporting this paper. Requests for further information or data should be sent to the authors.

\end{document}